# Deep Learning Enables Automatic Detection and Segmentation of Brain Metastases on Multi-Sequence MRI

**A PREPRINT**

[1,2]Endre Grøvik PhD[*], [3]Darvin Yi MS[*], [2]Michael Iv MD[†], [2]Elisabeth Tong MD[†], [3]Daniel L. Rubin MD MS[†], [2]Greg Zaharchuk MD PhD[†].

**ABSTRACT:** Detecting and segmenting brain metastases is a tedious and time-consuming task for many radiologists, particularly with the growing use of multi-sequence 3D imaging. This study demonstrate automated detection and segmentation of brain metastases on multi-sequence MRI using a deep learning approach based on a fully convolution neural network (CNN). In this retrospective study, a total of 156 patients with brain metastases from several primary cancers were included. Pre-therapy MR images (1.5T and 3T) included pre- and post-gadolinium T1-weighted 3D fast spin echo (CUBE), post-gadolinium T1-weighted 3D axial IR-prepped FSPGR (BRAVO), and 3D CUBE fluid attenuated inversion recovery (FLAIR). The ground truth was established by manual delineation by two experienced neuroradiologists. CNN training/development was performed using 100 and 5 patients, respectively, with a 2.5D network based on a GoogLeNet architecture. The results were evaluated in 51 patients, equally separated into those with few (1-3), multiple (4-10), and many (>10) lesions. Network performance was evaluated using precision, recall, Dice/F1 score, and ROC-curve statistics. For an optimal probability threshold, detection and segmentation performance was assessed on a per metastasis basis. Wilcoxon rank sum test was used to test the differences between patient subgroups. The area under the ROC-curve (AUC), averaged across all patients, was 0.98±0.04. The AUC in the subgroups was 0.99±0.01, 0.97±0.05, and 0.97±0.03 for patients having 1-3, 4-10, and >10 metastases, respectively. Using an average optimal probability threshold determined by the development set, precision, recall, and Dice-score were 0.79±0.20, 0.53±0.22, and 0.79±0.12, respectively. At the same probability threshold, the network showed an average false positive rate of 8.3/patient (no lesion-size limit) and 3.4/patient (10 mm$^3$ lesion size limit). In conclusion, a deep learning approach using multi-sequence MRI can aid in the detection and segmentation of brain metastases.

[*]*Co-first authorship (alphabetic order)*
[†]*These authors contributed equally to this work (alphabetic order)*

[1]Department of Radiology, Stanford University, Stanford, USA; [2]Department for Diagnostic Physics, Oslo University Hospital, Oslo, Norway; [3]Department of Biomedical Data Science, Stanford University, Stanford, USA

**Corresponding Author:** *Greg Zaharchuk, Department of Radiology, Stanford University, School of Medicine, 1201 Welch Road, Stanford, California 94305-5488, USA. Phone: (650) 736-6172, Fax: (650) 723-9222. Email: gregz@stanford.edu.*

## INTRODUCTION

Attributed in large by advances in effective and systemic treatment regimens of primary tumors, there has been an increase in the number of patients with metastatic cancer over the last decade (1). Brain metastases are one of the most common neurologic complications of cancer, most frequently originating from lung cancer, breast cancer, and malignant melanoma (2). In a survey including more than 26,000 patients, 12.1% of all patients with metastatic disease had brain metastases at diagnosis (3). Most patients present with three or fewer metastases to the brain, but 40% of patients have greater than this number (4, 5). Contrast-enhanced magnetic resonance imaging (MRI) is the key imaging technique in the diagnosis of brain metastases and is also used for longitudinal follow-up to assess treatment response.

The detection and segmentation of brain metastases is essential in the management of patients with brain metastases. Manual detection and segmentation of brain metastases is tedious and time-consuming, particularly with the growing use of multi-sequence 3D imaging, but it is an important task for many radiologists that must be performed with high accuracy. This is most clearly demonstrated in segmenting brain metastases for planning stereotactic radiosurgery (SRS), given that overinclusion will result in healthy tissue being irradiated and under-inclusion will lead to inadequate treatment. Furthermore, the diagnostic methods for assessing treatment response follow the criteria formulated by the Response Assessment in Neuro-Oncology (RANO) working group and are based on measuring the size of the enhancing lesion on Gadolinium (Gd) enhanced T1-weighted MR images (6). The traditional metrics used for response evaluation are based on unidimensional measurements, although the value of using volumetric measurements has been increasingly discussed. One concern raised by the RANO-group was that volumetric analysis, as performed manually by radiologists, adds cost and complexity and is not available at all centers.

During recent years, advances in machine learning (ML) have suggested the possibility of new paradigms in healthcare. One application of ML in radiology is the detection and segmentation of organs and pathology (7–10). In particular, there has been a significant effort in developing deep learning (DL) algorithms to learn from the comprehensive voxel-wise labeled MRI data for segmenting primary brain tumors (11–15). However, only a few studies have applied such ML approaches on patients with brain metastases (16–18), which may require different approaches given their size and multiplicity. To this end, this work demonstrates the use of a fully convolution neural network (CNN) for automatic detection and segmentation of brain metastases using multi-sequence MRI data as input.

## MATERIALS AND METHODS

### Patient Population and Imaging Parameters

This retrospective, single center study was approved by our Institutional Review Board. Inclusion criteria included the presence of known or possible metastatic disease (i.e., presence of a primary tumor), no prior surgical or radiation therapy, and the availability of all required MR imaging sequences (see below). Only patients with ≥ 1 metastatic lesion were included. Mild patient motion was not an exclusion criterion. To be most inclusive, we included cases performed at both 1.5T and 3T from all our clinical scanners, which included all major vendors. Based on these criteria, a consecutive set of patients were identified, imaged between June 2016 and June 2018. Details on this cohort are shown in Table 1, along with a list of the primary cancers.

The imaging protocol included pre- and post-gadolinium T1-weighted 3D fast spin echo (CUBE), post-gadolinium T1-weighted 3D axial IR-prepped FSPGR (BRAVO), and 3D CUBE fluid-attenuated inversion recovery (FLAIR). All sequences with key imaging parameters are summarized in Table 2.

### Image Segmentation and Co-registration

Ground truth segmentations were established by two experienced neuroradiologists by manually delineating and cross-checking regions of interest (ROI) around each enhancing metastatic lesion. The lesions were outlined on each slice on the post-gadolinium 3D T1-weighted IR-FSPGR sequence, with additional guidance from the 3D FLAIR and the post-gadolinium 3D T1-weighted spin echo data using the OsiriX MD software package (Version 8.0, Geneva, Switzerland).

Pre/post-contrast T1 CUBE and FLAIR images were co-registered to the IR-FSPGR space by normalized mutual information co-registration using the nordicICE software package (NordicNeuroLab, Bergen, Norway). Prior to network training, the brain was extracted by

**Table 1:** Demographics

| | |
|---|---|
| Total number of patients | 156 |
| Gender | 105 Female / 51 Male |
| **Primary cancer**: | |
| Lung | 99 (63%) |
| Breast | 33 (21%) |
| Skin/melanoma | 7 (5%) |
| Genitourinary | 7 (5%) |
| Gastrointestinal | 5 (3%) |
| Miscellaneous | 5 (3%) |
| **Number of Metastases:** | Number of patients |
| ≤3 | 64 (41%) |
| 4-10 | 47 (30%) |
| >10 | 45 (29%) |

**Table 2**: Overview of MRI pulse sequences and key imaging parameters

| Technique | 3D T1 BRAVO | Pre/Post 3D T1 CUBE | 3D CUBE FLAIR |
|---|---|---|---|
| TR (ms)[*] | 12.02 / 8.24 | 550 / 602 | 6000 |
| TE (ms)[*] | 5.05 / 3.24 | 9.54 / 12.72 | 119 / 136 |
| Flip angle[*] | 20 / 13 | 90 | 90 |
| FOV (mm$^2$) | 240×240 | 250×250 | 240×240 |
| Inversion time (ms)[*] | 300 / 400 | - | 1880 / 1700 |
| Acquisition matrix | 256×256 | 256×256 | 256×256 |
| Slice thickness (mm) | 1 | 1 | 1 – 1.6 |
| # of slices | 160 | 270 – 320 | 270 – 320 |
| Slice acquisition plane | Axial | Sagittal | Sagittal |

*TR = repetition time, TE = echo time, FOV = field-of-view, BRAVO – T1-weighted inversion recovery prepped fast spoiled gradient-echo, CUBE – T1-weigthed fast spin-echo, FLAIR – fluid attenuated inversion recovery.*
[*] *In case of varying parametric values between field strength, '/' notation is given (1.5T / 3T)*

using the Brain Extraction Tool (BET) (19) and applying the resulting brain masks on the network's input data. The brain masks were generated from the pre-contrast T1-weighted 3D CUBE imaging series and propagated to the other sequences.

**Convolutional Neural Network Details**
Training was performed using a 2.5D fully CNN based on the GoogLeNet architecture (20) (Figure 1). The network was modified to optimize segmentation by skipping the first and third down-sampling max pooling layers and using a stride of one on the first 7x7 convolutional layer. As a result, the final down-sampling rate throughout the convolutional layers were 4×, rather than 32×. To make the network fully convolutional, GoogLeNet's final fully connected layers were replaced by a single convolutional transpose layer of stride 4 and size 8x8. The final prediction was made on a single channel of logit values with a sigmoid cross-entropy loss function. In order to counter learning hurdles introduced by an unbalanced dataset, the loss on positive ground truth voxels were weighted 10× more than the loss on negative ground truth voxels.

To better capture through-plane features without incurring the inefficiencies associated with true 3D CNN's, we implemented a '2.5D' model. The network's input were 7 slices from each of the four aforementioned sequences, comprising a single center slice with 3 slices above and below, resulting in an input channel dimension of 28. Each image was rescaled (if necessary) to a size of 256×256. Note that all MR images were originally 256×256 or 512×512, and that bilinear interpolating the 512×512 images down to 256×256 gave minimal artifacts. Prior to training, pre-processing and normalization was performed with independent histogram equalization on each slice of the 28-channel input. During training, we randomly flipped and rotated the images in multiples of 90 degrees for data augmentation.

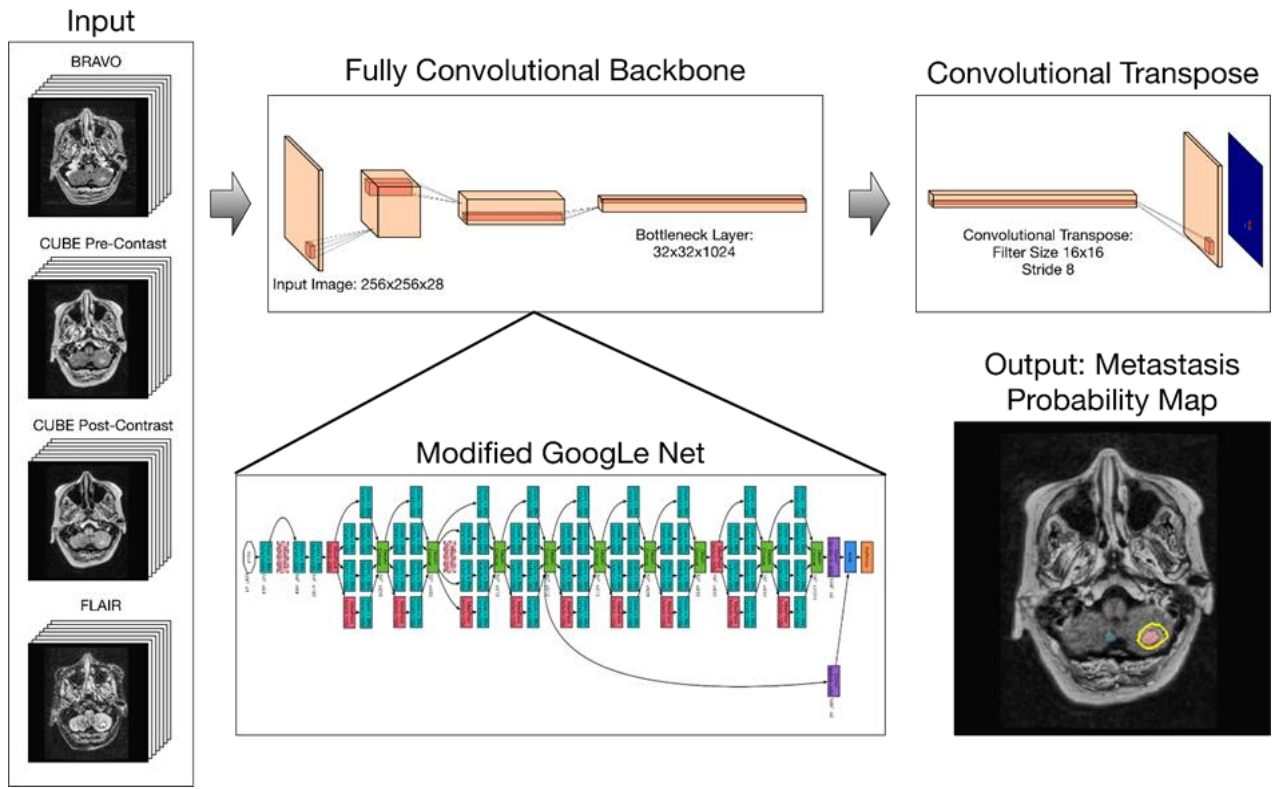

***Figure 1:*** *Flowchart showing the four image inputs used to train the neural network, the modified GoogLeNet architecture, and the resulting output colormap (overlaid a post-contrast BRAVO image) representing a probability map of whether the voxel represents a metastasis, ranging from 0-1.*

All training was performed on two consumer-grade graphical processing units (GPU's) (NVIDIA GeForce GTX 1080TI). The batch size was 32 with a learning rate of 0.001. Given that there were far more frames (>30×) without lesions compared to frames with lesions, we employed an uneven sampling procedure. For 16 of 32 images in each batch, we sampled the image from the set of frames with at least some lesion. For the other 16 images, we sampled uniform randomly from all frames. This ensured that for each batch, at least half of the images was populated with frames including some ground truth lesions. Regularization was performed by a L2 weight decay with a decay constant of $1e^{-5}$. Batch normalization was used following every convolutional layer. We used the ADAM optimization method (21) with default TensorFlow beta values of 0.9 and 0.999. By defining an epoch as the statistical equivalent of seeing every distinct frame of the dataset once, the training continues until convergence, which occurred at about the $10^{th}$ epoch. The network was trained using TensorFlow, and the resulting output was an image for each slide representing a probability map of whether the voxel represents a metastasis, ranging from 0-1.

The total number of cases were randomly broken into separate train, development, and test sets. None of the cases in the test set were present in the training set. To ensure a representative sample in our test set, we first chose the test cases as follows. First, we determined the number of distinct metastatic lesions in each case in the entire cohort and then broke the data into groups with (a) 1-3, (b) 4-10, (c) >10 lesions. We then randomly selected 17 cases from each of these groups, leading to a total test set size of 51 cases. The remaining cases were divided into training and development sets in a random 20:1 ratio.

**Statistical Analysis**

The network's ability to detect metastases on a voxel-by-voxel basis was evaluated using receiver operating characteristic (ROC) curve statistics, measuring the area under the ROC curve (AUC) for each patient in the test set. Only voxels within the brain mask were considered when calculating AUC. Corresponding sensitivity and specificity were determined by using the maximum value of the Youden's index as a criterion for selecting the optimal cut-off point. Based on ROC statistics from the development set, the optimal probability threshold for including a voxel as a metastasis was determined, and using this threshold, the results were further evaluated in terms of detection accuracy using precision and recall, and segmentation accuracy using the Dice similarity score (also known as the F1 score). In addition to the voxel-by-voxel analysis, the detection performance was also evaluated on a lesion-by-lesion basis by calculating the number of false positive (FP) per case. These metrics were determined by comparing the ground truth maps and the prediction maps, counting the number of overlapping objects using a connecting component approach. The number of FP were determined both without and with a lesion-size criterion, in which only objects ≥ 10 $mm^3$ were considered a detected lesion. Finally, the Wilcoxon rank sum test was used to compare the detection and segmentation metrics between the patient subgroups. A statistical significance level of 5% was used. All statistical analyses were performed using MATLAB R2017a version 9.2.0 (MathWorks Inc. Natick, MA).

**RESULTS**

A total of 156 patients were included in the study, with a final breakdown of 100 training cases, 5 development cases, and 51 test cases. The total time for training the neural network was approximately 15 hrs. For processing a test case, the forward pass on a single NVIDIA GTX 1080Ti GPU took less than 200 ms per slice during test time with a run time of approximately 1 minute for a full MR volume. Mean patient age was 63±12 years (range: 29-92 years). Primary malignancies included lung (n=99), breast (n=33), melanoma (n=7), genitourinary (n=7), gastrointestinal (n=5), and miscellaneous cancers (n=5). Of the 156 patients included, 64 (41%) had 1-3 metastases, 47 (30%) had 4-10 metastases, and 45 (29%) had >10 metastases. The test set of 51 patients had a total of 856 lesions.

Figure 2 shows an example case demonstrating the resulting probability map as an overlay on the post-gadolinium FSPGR image series using a lower probability threshold of 0.1. The voxel-by-voxel detection performance showed an area under the ROC-curve, averaged across all patients, of 0.98±0.04, corresponding to a sensitivity and specificity of 94% and 97%, respectively, at the optimal cut-off point. Further, the subgroups showed an area under the ROC-curve of 0.99±0.01 for patients having 1-3 metastases, 0.97±0.05 for 4-10 metastases, and 0.97±0.03 for >10 metastases, respectively (Fig. 3). The corresponding sensitivity and specificity are shown in Table 3A. The average optimal probability threshold for including a voxel as a metastasis, measured in the development set, was 0.93. Using this threshold, the precision, recall, and Dice-score were 0.79±0.20, 0.53±0.22, and 0.79±0.12, respectively. The distribution of these metrics within the subgroups is shown Table 3B. On a lesion-by-lesion basis, and by using the optimal probability threshold (average sensitivity = 83%), the network showed an average FP-rate of 8.3 (no size limit) and 3.4 (10 mm$^3$ size limit) lesions per case, with the highest sensitivity and lowest numbers of FP in patients with few metastases (Table 3C). The P-values, testing the differences in all detection and segmentation metrics between the subgroups, are shown in Table 4A-C. Examples in representative cases with different numbers of metastases are shown in Figure 4.

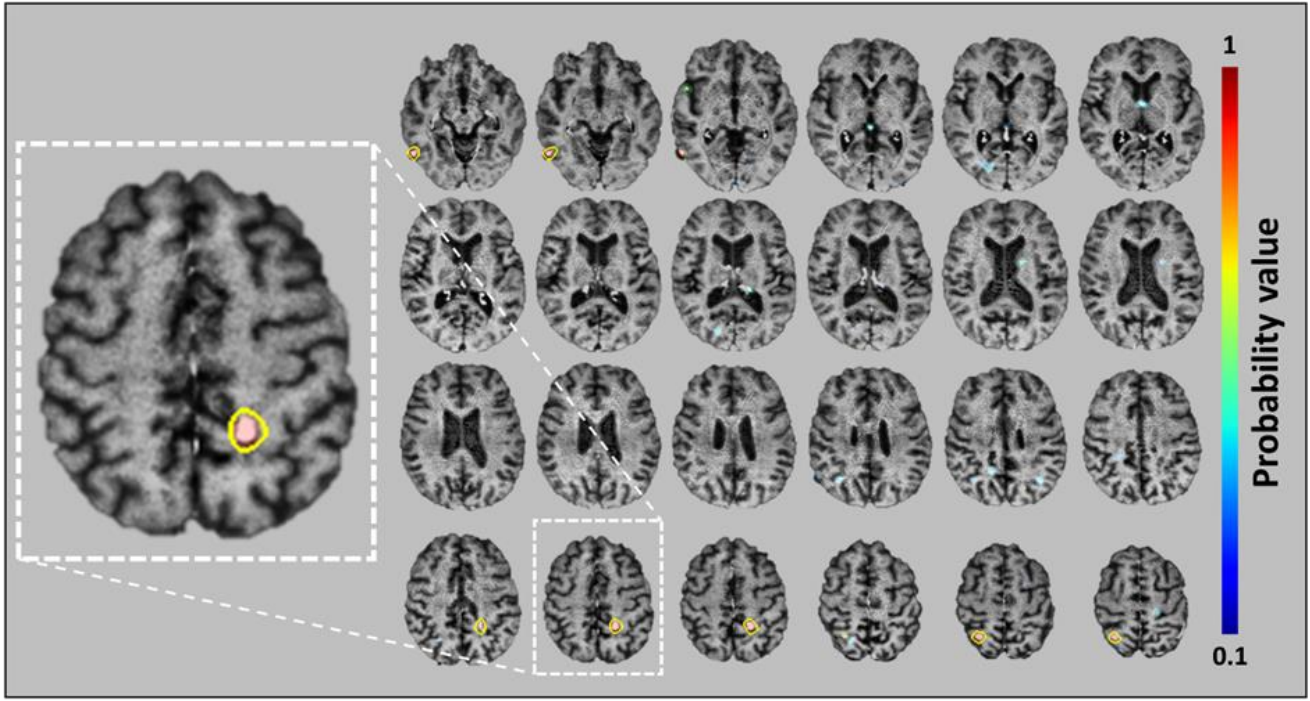

*Figure 2: Example case of a 47-year-old female patient presenting with three brain metastases from lung cancer. The image mosaic shows the predictions (probability maps as indicated by the color bar), generated by the neural network, and manually delineated metastases (yellow lines) overlaid the post-contrast image.*

### DISCUSSION

This study demonstrates that a modified 2.5D GoogLeNet CNN can detect and segment brain metastases on multi-sequence MRI with high accuracy. By testing on a large number of patients, thus facilitating subgroup analysis, this work demonstrates the network's clinical performance and potential, in addition to better understanding of its generalizability. To our knowledge, no previous study has reported on subgroup analysis using deep learning in brain metastases segmentation.

In recent years, many deep learning approaches have been developed and tested for automatic segmentation of gliomas (22), thanks in part to the publicly available BRAin Tumor Segmentation (BraTS) dataset. In contrast, only a few studies have used this approach for brain metastases. Liu et al. investigated the use of a CNN-based segmentation for SRS planning and reported a Dice-score of 0.67 and an AUC of 0.98 (16). The performance of the current method is superior to this prior method based on the average Dice-score (0.79 vs. 0.67), but showed similar AUC performance. Charron et al. used a 3D CNN (DeepMedic) for automatic detection

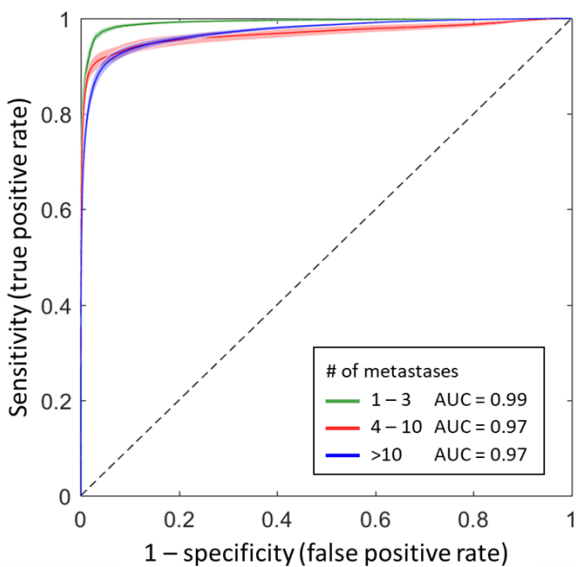

*Figure 3: ROC curves with 95% confidence intervals (shaded areas) for the three subgroups having 1-3 metastases (green), 4-10 metastases (red), and >10 metastases (blue). The average area under the ROC curve was 0.98, ranging from 0.79 to 1.00 for all cases.*

**Table 3: Summary of detection and segmentation metrics (mean value ± standard deviation)**

**A: Voxel-by-voxel detection accuracy using ROC statistics*** 

| # of metastases | AUC | Sensitivity | Specificity |
|---|---|---|---|
| 1 to 3 | 0.99±0.01 | 98±3% | 98±2% |
| 4 to 10 | 0.97±0.05 | 92±10% | 97±3% |
| >10 | 0.97±0.03 | 92±7% | 95±3% |
| All cases | 0.98±0.04 | 94±8% | 97±3% |

**B: Detection and segmentation accuracy at an optimal probability threshold****

| # of metastases | Dice | Recall | Precision |
|---|---|---|---|
| 1 to 3 | 0.76±0.20 | 0.54±.026 | 0.79±0.27 |
| 4 to 10 | 0.83±0.04 | 0.59±0.21 | 0.76±0.22 |
| >10 | 0.78±0.05 | 0.44±0.18 | 0.81±0.11 |
| All cases | 0.79±0.12 | 0.53±0.22 | 0.79±0.20 |

**C: Lesion-by-lesion detection accuracy at an optimal probability threshold****

| # of metastases | Sensitivity | FP (no size limit) | FP (10mm$^3$ size limit) |
|---|---|---|---|
| 1 to 3 | 92±25% | 3.2±4.0 | 1.7±2.0 |
| 4 to 10 | 81±19% | 8.5±9.8 | 4.4±6.0 |
| >10 | 76±20% | 13.1±18.9 | 4.1±10.3 |
| All cases | 83±22% | 8.3±12.9 | 3.4±7.0 |

AUC = area under the receiver operating characteristic (ROC) curve, FP = false positive.

* Sensitivity and specificity were determined by using the maximum value of the Youden's index

** The metrics were estimated using an optimal probability threshold of 0.93, as determined from the development set.

**Table 4: P-values comparing subgroups using Wilcoxon rank sum test**

**A: Voxel-by-voxel detection accuracy using ROC statistics***

| Subgroups | AUC | Sensitivity | Specificity |
|---|---|---|---|
| G1 vs G2 | **0.0131** | **0.0017** | **0.0496** |
| G1 vs G3 | **0.0024** | **0.0024** | **0.0038** |
| G2 vs G3 | 0.4282 | 0.6794 | **0.0421** |

**B: Detection and segmentation accuracy at an optimal probability threshold****

| Subgroups | Dice | Recall | Precision |
|---|---|---|---|
| G1 vs G2 | 1.0000 | 0.5816 | 0.2557 |
| G1 vs G3 | 0.0629 | 0.1131 | 0.2557 |
| G2 vs G3 | 0.0915 | **0.0230** | 0.8633 |

**C: Lesion-by-lesion detection accuracy at an optimal probability threshold****

| Subgroups | Sensitivity | FP (no size limit) | FP (10mm$^3$ size limit) |
|---|---|---|---|
| G1 vs G2 | **0.0069** | **0.0158** | 0.0829 |
| G1 vs G3 | **0.0002** | **0.0139** | 0.6352 |
| G2 vs G3 | 0.3952 | 0.7031 | 0.1178 |

G1 = subgroup having 1-3 metastases, G2 = subgroup having 4-10 metastases, G3 = subgroup having >10 metastases. Significant p-values are highlighted in bold. All P-values were measured using the Wilcoxon rank sum test.

* Sensitivity and specificity were determined by using the maximum value of the Youden's index.

** The metrics were estimated using an optimal probability threshold of 0.93, as determined from the development set.

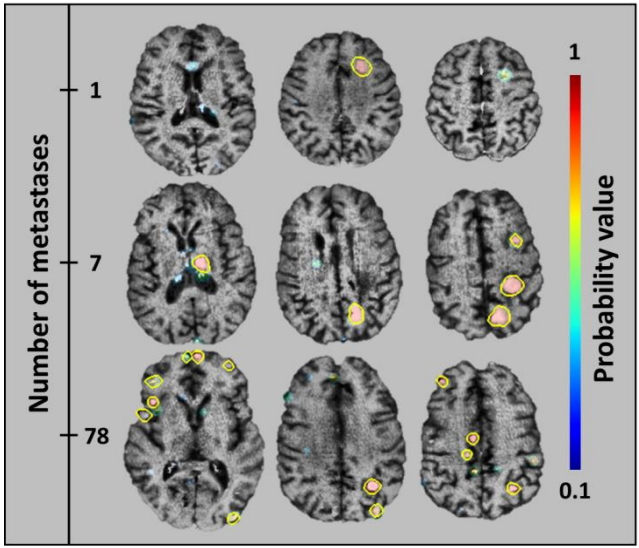

*Figure 4: Examples in representative cases with few (top row), moderate (middle row), and extensive (bottom row) metastatic disease. The top row shows a 70-year-old woman presenting with one brain metastasis from colon carcinoma. The middle row shows a 68-year-old man presenting with seven brain metastases from lung cancer. The bottom row shows a 37-year-old man presenting with seventy-eight brain metastases from lung cancer. The network's predictions are shown as probability maps and the yellow lines show the manually delineated lesions.*

and segmentation of brain metastases (17). By using segmented metastases to be irradiated as the ground truth, their network was trained using three MRI sequences from 146 patients as input and further tested on 18 patients. Similar to our study, their network was trained using three MRI sequences which proved to outperform networks trained on a single MRI contrast. Their network showed a sensitivity of 98% and 7.2 false positive per patient. Sunwoo et al. developed a computer-aided diagnostic (CAD) system for detecting brain metastases and a neural network for false-positive reduction (23). Their CAD system significantly improved the diagnostic performance of the reviewers and showed an overall sensitivity of 87%. One feature that separates our work from these previous studies is the strength of having diverse data, which may make it more challenging to demonstrate an overall high performance. We included cases from both 1.5T and 3T using multi-vendor scanners, and our data was not limited to patients receiving SRS, thus including more patients with extensive metastases disease (>10). This is supported by our results which indicate that the neural networks ability to detect and segment of brain metastases were reduced in patients with higher number of metastases. However, our patient cohort is more representative of real-world data. Furthermore, there are also differences in network architecture. Whereas Charron et al. used a full 3D network, we used a 2.5D network. The results indicate that our 2.5D network achieves the same segmentation performance as a 3D network, which reduces the computational and memory requirements for training. However, further studies systematically comparing 2D, 2.5D, and 3D neural architectures must be performed to adequately answer this question. Also note that, using comparable graphic cards, our 2.5D network required approximately 1 minute to perform a forward pass (inference) while a 3D network would require approximately 20 minutes for the same task.

The split between training/validation/testing in this study is somewhat unusual compared to similar studies in the literature. However, we chose to test on a large number of cases to understand how generalizable the network was and to facilitate subgroup analysis, enabling a better understanding of the network's clinical performance and potential. In earlier stages of the study, we found that the network had high performance training on approximately half of the current training set, and that increasing the number of training cases did not provide significant improvement, thus justifying our use of a larger test set. Our results indicate that the networks ability to detect metastatic voxels, as measured by the AUC, is best in patients with few (1-3) metastases and further that the segmentation performance, as measured by the Dice-score, is slightly better for patients with 4-10 metastases. On a lesion-by-lesion basis, our results suggest that the network performs best on patients with few metastases, both in terms of sensitivity and the number of false positive. One hypothesis is that these results may be associated with an optimal tradeoff between total number and individual size of the metastases.

Multiple network architectures were considered for this project, including residual networks (24), dense networks (25), U-Nets (26), Pyramid Scene Parsing (PSP) Nets (27), and Feature Pyramid Networks (FPNs) (28). However, after running preliminary experiments with the 2014 GoogLe (or Inception v1) network, we found that this was already capable of overfitting the training data. Thus, given that network complexity and capacity were not driving issues in the project, and that the GoogLeNet enables high computational efficiency, both in memory and speed, this became our choice of architecture. The compact size of the network allows it to be run on even the smallest mobile GPUs, such as the NVIDIA Tegra chip. However, note that more ample gains from advanced network architectures could be attained with larger datasets.

Typical workflow in radiotherapy planning requires accurate detection by a radiologist, followed by segmentation by a radiation therapist. Both steps are time-consuming and subject to interobserver variation. Detection requires manual visualization and annotation. Fatigue and image quality are a few factors that may affect the accuracy (29, 30). Special imaging techniques have been proposed to improve this process. For instance, double dose gadolinium-based contrast-enhanced thin-slice MRI produced more precise delineation of lesions compared to using single dose (31). The addition of overlapping CUBE maximum intensity projection (MIP) images, which have better contrast-to-noise ratio of metastatic lesions than post-gadolinium 3D isometric high-resolution sequences, are often used to enhance the sensitivity of detection. However, even with the addition of CUBE MIP images, the interrater agreement for identification of metastases between 2 experienced radiologists was reported as only fair-to-moderate in one study (32). Segmentation requires tracing the contours of the lesions on the 2D images slice by slice. Even though there is semi-automatic software available for segmentation, extensive manual editing is often required, thus generating non-reproducible operator-dependent results (33, 34). Accurate segmentation of the metastases is imperative in radiation therapy planning to minimize damage to adjacent normal tissue. Our neural network essentially combines visualization, quantification, and segmentation into one fluent step, producing results that can be directly applied to radiotherapy planning, with minimal user-interaction.

While this study shows high accuracy and performance using deep learning for segmenting brain metastasis, several potential study limitations exist. First, the results must be interpreted in light of the limited sample size in this single center, retrospective study. This is partly related to the time required for manual segmentation. Future studies will investigate the use of 'coarse' segmentation, which is by far less time-consuming compared to fine segmentation, and how this may affect the network's ability to detect and segment brain metastases. Also, testing of the network performance on multi-site data remains a key step towards understanding its clinical value. Second, the network sometimes fails in terms of reporting FP. This is particularly true in and near vascular structures at the skull base such as venous sinuses, or over the cortex. Finally, as our neural network is trained on four distinct MRI contrasts, the use of this method is limited to sites acquiring all sequences. However, future studies will address the issue of having other or even lacking model inputs with the aim of making the neural network more robust and versatile towards different input channels.

In conclusion, our study demonstrates that a deep learning network can detect and segment brain metastases on multi-sequence MRI with high accuracy and thus illustrates the potential use of this technique in a clinically relevant setting.